# Induced Optical Losses in Optoelectronic Devices due to Focused Ion Beam Damages

F. Vallini[1], L. A. M. Barea[1], E. F. dos Reis[1], A. A. von Zuben[1] and N. C. Frateschi[1]

[1]All authors from Laboratório de Pesquisa em Dispositivos, Departamento de Física Aplicada, Instituto de Física Gleb Wataghin, Universidade Estadual de Campinas, Campinas, SP, Brasil
e-mail: fvallini@ifi.unicamp.br

*Abstract*— **A study of damages caused by gallium focused ion beam (FIB) into III-V compounds is presented. Potential damages caused by local heating, ion implantation, and selective sputtering are presented. Preliminary analysis shows that local heating is negligible. Gallium implantation is shown to occur over areas tens of nanometers thick. Gallium accumulation as well as selective sputtering during III-V compounds milling is expected. Particularly, for GaAs, this effect leads to gallium segregation and formation of metallic clusters. Microdisk resonators were fabricated using FIB milling with different emission currents to analyze these effects on a device. It is shown that for higher emission current, thus higher implantation doses, the cavity quality factor rapidly decreases due to optical scattering losses induced by implanted gallium atoms.**

*Index Terms*—**Focused Ion Beam, Implantation and Microdisk resonators.**

## I. INTRODUCTION

Focused Ion Beam (FIB) is a promising tool for nanodevices fabrication. FIB allows milling with very good morphology and anisotropy, and precise deposition of metals and dielectrics, all with nanometric precision. Moreover, FIB processes can easily be inserted in the chain of conventional microfabrication of electronic or optical devices. Several works have shown the fabrication of devices, with very low roughness, in very complex structures as light emitting resonators, based on Fabry-Perot or microdisk architectures (1, 2). Nevertheless, there are still some doubts whether high performance electronic or optoelectronics devices can be obtained using this technique due to intrinsic damages caused by FIB. In the case of optoelectronic devices, it is important to investigate the damages caused by milling of III – V materials. This work presents a study of the damages due to local heating, ion implantation, and selective sputtering on GaAs and InP. Also, we investigate spectral properties of microdisk resonators fabricated by FIB under different milling conditions to inspect induced optical losses. These optical losses have two main causes: the increase of photon absorption due to changes in the instrinsic properties of the crystal and increase in linear optical scattering due to morphological changes and the presence of a larger number of scattering centers after milling and, hence, ion implantation. All fabrication in this work was performed in a FEI NOVA 200 dual beam system FIB/SEM. Characterization of these devices confirms that ion implantation creates scattering centers that increase the optical losses of the resonators due to linear optical scattering losses.

## II. LOCAL HEATING AND ION IMPLANTATION

Several works present thermal processes for the healing of possible structural damages caused by ion beam milling [1,2]. Nevertheless, addressing some of the causes of the damages may help optimizing processes and reduce these damages. Two important effects are investigated: heating and ion implantation during fabrication.

### A. Local Heating

Local heating is generated by energy transfer from the ion beam to the sample during milling. Assuming all the beam energy is transferred to the substrate as heat, the heating power is given by the product of accelerating voltage $V$ and emission current $I$. This is a worst case scenario, since great part of the energy is used to remove atoms from the lattice and perform the milling.

In order to evaluate the local heating, we use finite element method based software, COMSOL, to solve the heat transfer equation:

$$-\nabla.(k\nabla T) = C(T_0^4 - T^4) \quad (1)$$

Where k is thermal conductivity of the sample, C is the Stefan-Boltzmann constant, $T_0$ is the chamber temperature (284 K) and T is the temperature scalar field. Assuming the beam will impinge perpendicularly to the sample surface, and that $q_0$ is the heat flux inside this surface, the boundary condition is given by

$$\vec{n}.(k\nabla T) = q_0 + C(T_0^4 - T^4) \quad (2)$$

In order to calculate the heat flux we need to obtain the beam intensity. For each emission current $I$ there is a different beam area, $A(I)$. These values are provided by the FIB system manufacturer. Thus, the heat flux is given by $q_0 = VI/A(I)$. To have an idea of the typical values of local heating, for an accelerating voltage of 30 kV and an emission current of 0.5 nA, the beam diameter is about 39 nm and the beam energy flux is $1.24 \times 10^{11}$ W/m$^2$. With the heat flux, equation (1) can be solved to give the steady state temperature field. The local heating is given by the difference between the maximum temperature, located at the beam position in the substrate, and a heat sink placed 10 microns away. The heat sink is at the same temperature as the chamber. The local heating for a wide range of emission currents is shown in figure 1. The substrate simulated is InP with a heat conductivity of $k_{InP} = 0.68$ Wcm$^{-1}$°C$^{-1}$. The range of emission currents used in this plot is much wider than typically used for fabrication.

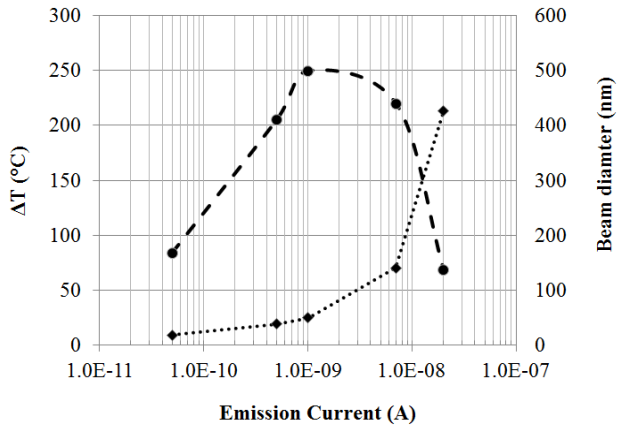

Fig. 1: Dash - left: Local heating for different beam current emissions. Dot - right: Beam diameter for different current emission (FEI NOVA 200 dual beam system FIB/SEM)

Figure 1 shows that the maximum local heating is about 260 °C. This value is well within a safe temperature range for typical semiconductor processing and, therefore, no strong thermal damage is expected.

The fact that there is a maximum local heating is surprising. This is caused by fast increase of the beam diameter with the emission current. The dotted line in figure 1 shows the beam diameter values as a function of emission current for the FEI NOVA 200.

*B. Gallium implantation*

Since milling with FIB is realized with ions of Ga$^+$, it is expected that Gallium will be implanted on the sample. The implantation of an element can result in effects as doping and sometimes as quantum well intermixing, due composition alteration at the quantum well region, for example [3].

During the milling process, the ion beam hits the sample removing material and implanting ions in two main directions: in depth (along the ion beam direction) and transversely (perpendicular to the ion beam direction). Also, there is re-deposition of scattered ions during milling. To estimate how many ions are implanted during this process we realized a Monte Carlo simulation using the software The Stopping and Ranging of Ions on Matter (SRIM) [4]. To perform this simulation we considered 30 kV of ion acceleration energy and an emission current of 0.5 nA during 1 µs. The acceleration voltage and the emission current values have been chosen since they provide the best morphology for our devices. Also, 1 µs is the typical dwell time of the beam. This exposure corresponds to ~3x10$^3$ gallium atoms impinging the sample. The calculated Ga implantation depth is 23 nm. The lateral spread of the Ga implantation has a mean value of 30 nm and there is no implantation further than 70 nm from where the beam hits. These values indicate an approximately maximum implanted Gallium density of 2x10$^{21}$ cm$^{-3}$. The implantation profile should not depend on the emission current which only changes the implantation doses. Therefore, although there may be a large concentration of implanted ions, the implanted region will be very small after milling. To improve the performance of devices, one should develop diluted etching solutions to remove the thin layer with implanted gallium.

To experimentally investigate Gallium ion implantation, Electron Dispersive X-Ray (EDX) spectroscopy was performed on a GaAs sample before and after FIB milling. The milling was done to produce a rectangular hole of 10 x 7.5 µm with a depth of 0.5 µm. The accelerating voltage and the emission current was, respectively, 30 kV and 0.5 nA. The milling time was 30 s. Scanning electron microscope micrograph of the rectangular hole obtained for GaAs is shown in figure 2 (a). One observes a rough morphology with several clusters. We believe these clusters appear from segregation of the excess Gallium. The Gallium in excess is a result of implantation and/or arsenic species selective sputtering. EDX results of a region at the bottom of the rectangular hole are shown in table I. A 5% enhancement (reduction) is observed for gallium (arsenic). As mentioned above, the maximum implanted gallium density is estimated to be of the order of 2x10$^{21}$ cm$^{-3}$. The atomic density of GaAs is 4.42x10$^{22}$ cm$^{-3}$. Therefore the implanted Gallium increase should be of the order of 4.5% in good agreement with the EDX result. Therefore, we don't expect selective sputtering of Arsenic during the milling process. To further investigate this hypothesis we performed the same milling on an InP substrate. The binding energy for Phosphorus in InP is similar to that for Arsenic in GaAs. A micrograph of the InP sample after milling is shown in figure 2 (b). The morphology is much better, and the presence of clusters is not evident as in figure 2 (a). If Phosphorus is selectively sputtered from the sample, Indium atoms would migrate and form clusters similar to those observed for GaAs, but this phenomena is not observed. This process of cluster formation is well known for III-V compounds after group V species losses. For more complex III-V alloys, we expect a change in stoichiometry in the thin implanted region.

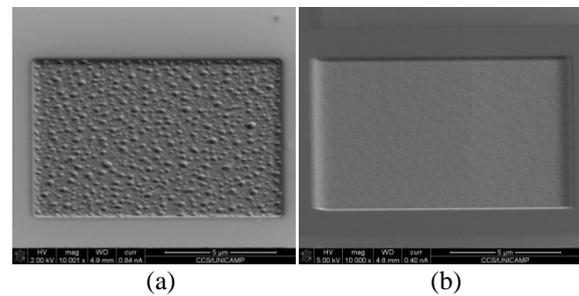

Figure 2. Rectangular hole 10 x 7.5 µm with depth of 0.5 µm milled with FIB with V = 30 kV and I = 0.5 nA for 30 s. (a) GaAs sample; (b) InP sample.

Table I. EDX Analysis of GaAs Before and After FIB

| Element | After FIB – Before FIB |
|---|---|
| Ga | + 4.9 % |
| As | - 5.0 % |

# III. FABRICATION AND CHARACTERIZATION OF MICRODISK RESONATORS

In order to evaluate the FIB induced damages directly, we choose to fabricate a microdisk resonator device, whose optical confined modes are strongly affected by the sidewall roughness and the presence of scatering centers resulted from milling [5]. We fabricated three microdisk resonators with FIB, each one with a different ion beam emission current.

## A. Fabrication

The epitaxial layers, shown on the inset of figure 3 (a), are grown on a n-type (001) InP substrate on the sequence: Si-doped ($5 \times 10^{17}$ cm$^{-3}$) InP lower cladding layer, undoped InGaAsP multiquantum well active region within a 300-nm InGaAsP waveguide layer, followed by a 1.7 µm thick Zn-doped top p-type cladding layer. In this layer the doping concentration increased from $2x10^{17}$ cm$^{-3}$ to $5x10^{18}$ cm$^{-3}$ with the growth. A highly Zn-doped p-InGaAs top contact layer of 200 nm was employed for p-contact [6]. First, Au/Ge/Ni is deposited over all the substrate to do the n ohmic contact. The sample was alloyed in forming gas for 30 s at 420° C. On the p-side of the sample Ti/Pt/Au microdisk structures were evaporated by electron-beam evaporation followed by conventional lift-off. Subsequently the milling was done with Ga$^+$ ions using a dual-beam FIB/SEM (focused-ion beam/scanning electron microscope). The milling condition was: emission currents of 7 nA, 0.5 nA and 0.1 nA with milling times of 50 s, 697 s and 1560 s, respectively, all with 30 kV to reach a depth of 6 µm, which was well within the InP substrate. In this step, we removed the field around the microdisk, obtaining several pillars, all with the top metallization format. These pillars have 8 µm radius that will be the microdisk radius. As result, we obtained pillars with very smooth walls. The final step consists of a wet-chemical etching with H$_3$PO$_4$ and HCl to selectively remove InP material, leaving suspended disks structures of the InGaAs contact layer and the InGaAsP active region/waveguide. A finished microdisk resonator is shown in figure 3 (b).

## B. Characterization

The resonators were characterized by electroluminescence with the set-up shown in figure 3 (a). The devices were mounted on a Peltier cooled stage set at 20 °C. Current injection was done using a force sense system and a microprobe. The spectra for the three microdisks were measured using an Optical Spectrum Analyzer (OSA). The light emitted from the microdisk was collected using an optical fiber with a GRIN rod lens collimator followed by micro-lens with 300 µm focal length. Figure 4 shows the normalized spectra for the three disks (a) FIB current of 0.1 nA; (b) FIB current of 0.5 nA; and (c) FIB current of 7.0 nA. All devices were pumped with a current injection of 10 mA. The OSA is set to 1 nm resolution bandwidth and 10 times average.

The spectra show the typical whispering gallery resonance modes [5]. The modal separation for the three spectra is between 14 nm and 18 nm. For 8 µm radius disk, the calculated modal separation is 17 nm for an effective index of 2.5. It is important to observe that the resonances are less pronounced for the highest FIB current case. A simple calculation of the quality factor for the main peaks (ratio between the line width, FWHM, and the center wavelength, all above the emission background) shows the spectral deterioration more quantitatively. The quality factor is of the order of 240 and 200 for disks (a) and (b), respectively, while it is about 130 for the disk (c). Since the whispering gallery modes are located very close to the edge of the disk, the implantation that occurs near the edge of the disk during milling becomes very important because it introduces morphological defects and introduces optical scattering centers at the same location as the optical field maximum. The dependence of the implantation doses is always reduced to the last ion beam exposure for the removal of material. Therefore it is fixed by the dwelling time. Since the penetration range of the implantation is independent of the doses, we can assume that it occurs roughly in a 30 nm layer at the disk's edge. However, the number of implanted ions is directly proportional to the emission current, being $4.4x10^{20}$ cm$^{-3}$, $2.2x10^{21}$ cm$^{-3}$, and $3.1x10^{22}$ cm$^{-3}$ for the disks (a), (b), and (c), respectively.

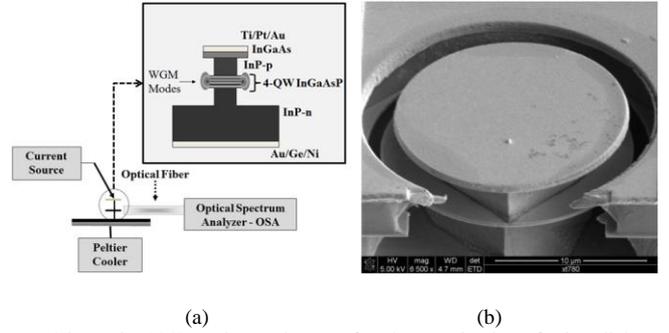

Figure 3. (a) Experimental set-up for characterization of microdisk resonators. The inset is the schematics of a fabricated microdisk resonator with the epitaxial struture. (b) Micrograph of the microdisk resonator of 8 um radius fabricated with FIB.

An important parameter to be investigated is the acceleration voltage of FIB, as it determines the depth of implantation. Reducing the acceleration voltage causes a reduction in the implantation depth during the dwell time. However, for the same beam power required for a giving milling rate, higher undesired implantation doses will result. A detailed study including the effects of the acceleration voltage is ongoing.

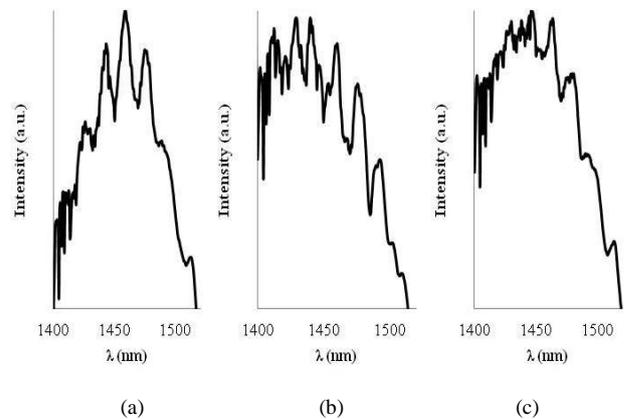

Figure 4. Emission spectrum of 8 µm radius microdisk resonators pumped with a current of 10 mA at 20 ºC. (a) FIB current of 0.1 nA; (b) FIB current of 0.5 nA; and (c) FIB current of 7.0 nA.

## C. Induced optical losses analyzes

We expect that all damages related to implantation should be proportional to the implantation doses. For instance, vacancies, amorphization, presence of interstitial gallium and even stoichiometry change and intermixing of the quantum well/barrier material should be more pronounced for higher doses. Since optical gain is provided by the quantum wells, this can result in high absorption and the reduction of Q.

To better understand the reason for the Q reduction, our approach is to calculate the overlap between the optical confined modes at the microdisk resonator with the active region implanted with Gallium and see how the losses are affected when increasing the implantation density.

It is well known that the modes in a cylindrical dielectric disk cavity with radius much larger than the wavelength can be approximated by [5]:

$$\Psi_{in} = A_{m,n} J_m\left(\frac{n_{eff}\, \omega_{m,n} r}{c}\right) e^{im\varphi} \quad (3)$$

where A is a normalization constant, J is a Bessel function of azimuthal order m, r is the radial position, $n_{eff}$ is the effective index of the dielectric cavity for the axial confinement, ω is the resonant frequency, c is the speed of light and φ is the azimuthal dependence of the field. n is the radial number equal 1 for the so called whispering gallery modes (WGM). The WGM frequencies for a microdisk with radius R are related with the first zero (n = 1) of the $m^{th}$ order Bessel function, $x_m$, by:

$$\omega_{m,1} = \frac{x_m c}{n_{eff} R} \quad (4)$$

In order to calculate the radial position of the measured modes it is not necessary to take care of the exponential component of the field, because it has spatial symmetry along φ-direction. So, we just need to compute the $m^{th}$ order Bessel function for each respective resonance frequency. For our microdisks, with radius 8 μm and effective index of 2.5, we compute modes from M = 83 to M = 89, with radial confined field shown on figure 5.

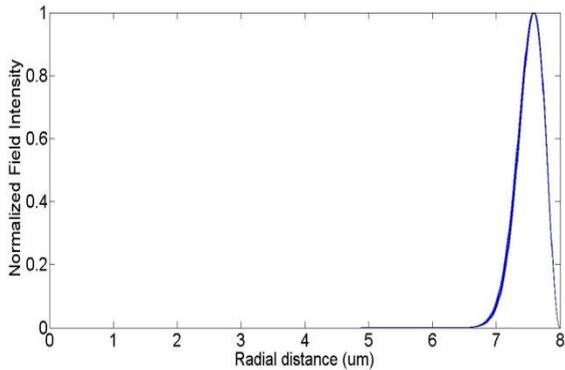

Figure 5: Normalized optical field intensity confined inside the microdisk resonator along the radial direction for modes M = 83 to M = 89.

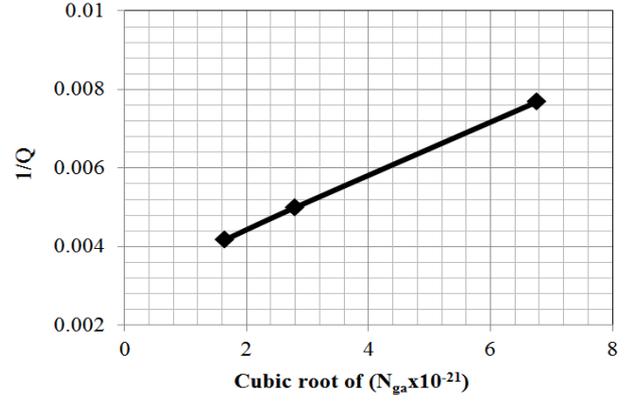

Figure 6: Inverse of quality factor dependence with the cubic root of the implanted Ga density.

As shown on figure 5, all the modes are spatially near each other, as expected for higher order WGM. The field intensity is very low at the 30 nm end of the disk, so there is a very small overlap between the optical field and the lateral region of implanted Gallium (~30 nm). Approximately, only 0.03 % of the optical mode interacts with the implanted region. Although the interaction is small, the small mode volume and the small cavity size can be severely affected by propagation losses.

In order to qualitatively understand the propagation loss mechanism that the implanted ions are inducing in the optical field we consider two possibilities: (1) Gallium ions lead to barrier-well intermixing and, thus, well-barrier composition change; (2) Implanted gallium species are acting only as scattering centers. In the first case, linear losses should depend exponentially with the linear density of implanted ions because the intermixing leads to a gain peak shift. On the other hand, scattering losses should scale linearly with the linear density of implanted ions.

The quality factor Q is inversely proportional to the cavity losses. The linear density of implanted ions is proportional to the cubic root of the implanted doses. Figure 6 shows the measured value of 1/Q versus the cubic root of the estimated implanted doses for disks (a), (b), and (c). A clear linear behavior is obtained. Therefore, we believe that the damages induced by the implanted ions are predominantly acting as scattering centers.

## IV. CONCLUSION

Three main possible problems of using FIB for III – V materials were explored. The local heating during milling was shown negligible but ion implantation has been shown to be important for devices with active material near the milled surface since most of the implantation occurs in a thin layer of tens of nanometers. Selective sputtering is not evident. In the case of GaAs, ion milling presents a serious problem of generating free excess gallium atoms on the surface that migrate to form metallic clusters. To analyze the ions effect on optoelectronic devices, microdisks resonators were fabricated with FIB showing that higher ion doses reduces the cavity Q-factor. The decrease in the quality factor is due to Ga implantation, but there are two factors: the excess of Ga and the implantation damage induced in the disk border. Finally we show that the implanted ions are predominantly acting as scattering centers, being responsible for induced optical losses.

## ACKNOWLEDGEMENTS

The authors would like to thank the financial agencies Fundação de Amparo à Pesquisa no Estado de São Paulo (FAPESP), Conselho Nacional de Pesquisa (CNPq), Coordenação de Aperfeiçoamento de Pessoal de Nível Superior (CAPES) and to the Centro de Componentes Semicondutores (CCS). This work was performed under the Instituto Nacional de Ciência e Tecnologia FOTONICOM.